# Spontaneous assembly of condensate networks during the demixing of structured fluids

Yuma Morimitsu[1]†, Christopher A. Browne[1]†, Zhe Liu[1], Paul G. Severino[2], Manesh Gopinadhan[3], Eric B. Sirota[3], Ozcan Altintas[3], Kazem V. Edmond[3], Chinedum O. Osuji[1]*.

[1] Department of Chemical and Biomolecular Engineering, University of Pennsylvania, Philadelphia PA 19104.
[2] Department of Physics and Astronomy, University of Pennsylvania, Philadelphia PA 19104.
[3] Research Division, ExxonMobil Technology and Engineering Company, Annandale, NJ 08801, USA

† Equal contribution authors. * Corresponding author: cosuji@seas.upenn.edu.

**Abstract.** Liquid-liquid phase separation, whereby two liquids spontaneously demix, is ubiquitous in industrial, environmental, and biological processes. While isotropic fluids are known to condense into spherical droplets in the binodal region, these dynamics are poorly understood for structured fluids. Here, we report the novel observation of condensate *networks*, which spontaneously assemble during the demixing of a mesogen from a solvent. Condensing mesogens form rapidly-elongating filaments, rather than spheres, to relieve distortion of an internal smectic mesophase. As filaments densify, they collapse into bulged discs, lowering the elastic free energy. Additional distortion is relieved by retraction of filaments into the discs, which are straightened under tension to form a ramified network. Understanding and controlling these dynamics may provide new avenues to direct pattern formation or template materials.

## Introduction

Liquid-liquid phase separation plays a key role in a range of industrial purification processes [1, 2], material fabrication pathways [3, 4], and biological compartmentalization of cellular processes [5-7]. Surface tension between the two liquid phases generally results in spherical condensate drops, which grow and coalesce. However, many relevant industrial and biological systems are composed of *structured* fluids—e.g. mesogens [8], lipids [9], colloids [10], chromatins [11], and many proteins [12]. Such fluids can exhibit "liquid crystallinity", in which molecules are ordered into regular mesophases, but nevertheless deform fluidly to external stresses [13, 14]. The processing of many industrial soft materials, optics, and consumer products involves the phase separation of liquid crystal mixtures [15-17], though it remains poorly understood whether liquid crystallinity can affect the dynamic shape of condensates during phase separation.

For many liquid crystals, e.g. nematics and cholesterics, the dominance of surface stress drives the formation of spherical interfaces, distorting the mesophase within the droplet [18-22]. In contrast, smectic liquid crystals have much larger elastic stresses associated with bulk mesophase distortion that can compete with surface stresses [23-26]. As a result, unusual filamentous domains can form during the isotropic-smectic mesophase transition for liquid crystal mixtures [27-29], driven by the distortion energy of the smectic layering [30]. These filaments rapidly elongate due to a preferred growth direction, driving strong local hydrodynamic flows that buckle the filaments under viscous drag [31]. Similar filamentous structures can form during the simultaneous demixing of smectic-A-forming mesogens from isotropic solvents [32-36], suggesting a broader unexplored potential for liquid crystallinity to alter the dynamics of phase separation and reshape condensates.

Here, we provide the first report of a striking new condensate geometry accessible by demixing. Upon cooling, a mesogen-enriched condensate dynamically self-assembles into a complex porous network. Optical microscopy reveals that this hierarchical architecture is formed by the growth of filaments that intermittently collapse into discs, mediated by a competition of hydrodynamics and relief of elastic mesophase distortion. The resulting networks form in minutes but remain "living" for up to an hour during sustained cooling—rearranging continuously as new filaments form, traverse the pore space, and form new network connections. Varying the cooling rate provides one knob to tune the network architecture. Our work demonstrates that liquid crystallinity can dramatically reshape condensates into patterned hierarchical architectures. Given the ubiquity of liquid crystals in industrial and biological fluids, understanding and controlling these dynamics may provide new insights into spontaneous pattern formation in nature and new avenues for the self-assembly or templating of materials.

**Condensing liquid crystals form sparse network architecture**

We study the thermally-induced phase separation of a model liquid crystal mesogen from an isotropic solvent. Our mesogen is 4'-cyano 4-dodecyloxybiphenyl, commonly referred to as 12OCB, which forms a smectic A liquid crystal in isolation, and is dissolved at all compositions in our solvent is squalane at elevated temperatures >100°C (Supplemental Materials section 1). Differential scanning calorimetry (Fig. S1) and X-Ray scattering (Fig. S2) are used to characterize the phase space (Fig. S3), which exhibits a two-phase region in which an isotropic solution and a smectic A liquid crystal phase coexist. To investigate the dynamics of the demixing in this two-phase region, we use an intermediate composition of 0.45 weight fraction mesogen. We load this solution into a Hele Shaw imaging cell above $T_c \approx 66°C$, where the solution exists as a homogeneous isotropic liquid. We then slowly cool the solution at a rate of 0.1°C/min and image the dynamics of the demixing to form an isotropic and liquid crystal phase using brightfield (BF) and polarized optical microscopy (POM) (Figure 1A).

Near $T_c$, the condensing smectic-A phase spontaneously organizes into a dense filamentous structure, which expands as the filaments rapidly elongate and buckle (false-colored blue, Figure 1B, Movie S1). The condensing phase spontaneously organizes into a dense filamentous structure, which expands as the filaments rapidly elongate and buckle (false-colored blue, Figure 1B, Movie S1). Successive buckling forms a radial front, which disperses across the field of view (red dashed line). In the wake of this front, dense regions of filaments sporadically collapse to form irregular aggregates structures (false-colored yellow, purple circles indicate newly formed aggregates), which rapidly retract filaments. After an initially rapid re-organization, aggregates stabilize to form a sparse network tethered by straightened filaments under tension. This condensate network remains sparse and open under tension, presumably from a combination of continued filamentous dispersive growth at the periphery and friction with the top and bottom glass boundaries. After the initial network forms, new filaments nucleate and explore the pore space (red arrows), which continually reshape the network through the addition of aggregate nodes in their wake (purple circles). We use polarized optical microscopy (POM) and theory to understand the formation and dynamic evolution of this condensate network.

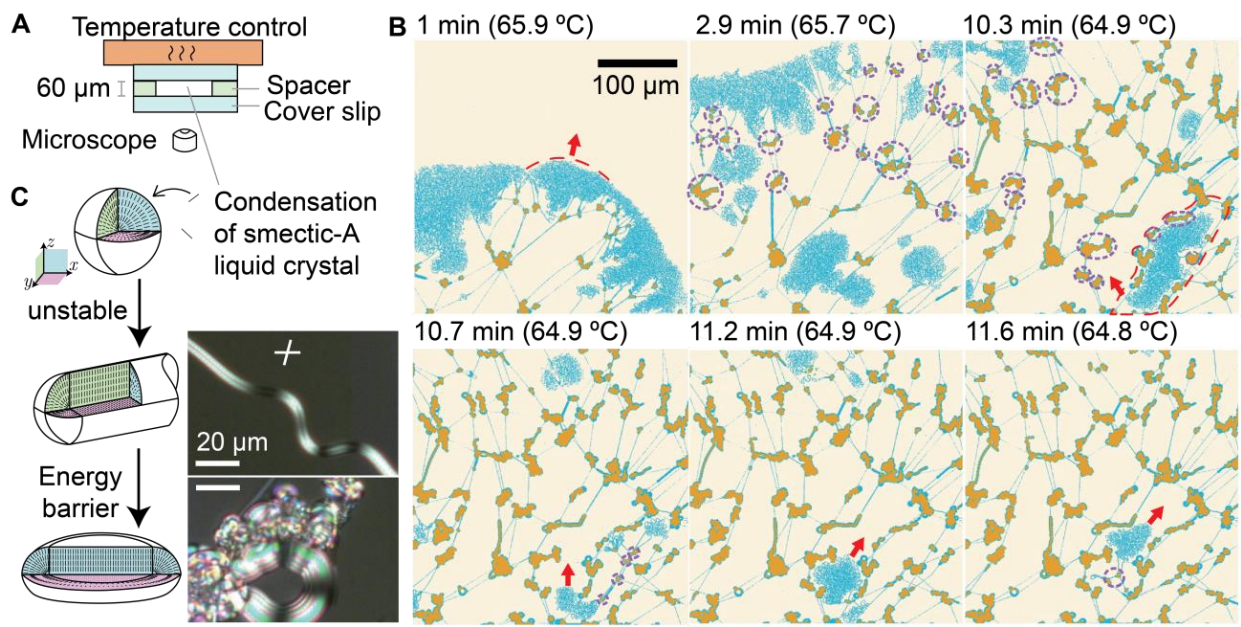

**Figure 1: Phase separation of liquid crystal forms nonequilibrium hierarchical network of filamentous and disc-like condensates. A** A homogeneous isotropic liquid composed of 0.45 weight fraction of a mesogenic liquid crystal (12OCB) and an isotropic solvent (squalane) is loaded into a 60 μm tall Hele-Shaw cell. The sample is slowly cooled from high temperature into the binodal region at −0.1°C (Supplemental Materials section 1). **B** Brightfield microscopy (BF) indicates the formation of condensates when subcooling beneath $T_c$, visualized by scattering associated with a refractive index difference. The condensing phase spontaneously forms rapidly-elongating filaments (false-colored blue using image thresholding), which expand as a front (red dashed lines and arrows) and intermittently collapse (purple dashed circles) into dense aggregates (false-colored yellow), which form a sparse porous network. Unprocessed images in Figure S5 and Movie S1. **C** Polarized optical microscopy (POM) indicates the condensing phase has liquid crystalline ordering, as schematized from POM analysis on left (Supplemental Materials section 1). Filaments exhibit radial splay. Aggregates are comprised of bulged discs, which have undistorted interiors and radial splay in the bulged periphery.

**Mesophase alignment drives condensation into filaments**

The initial formation of filamentous condensates, as characterized previously [32-35], is driven by the elastic energy penalty associated with the splay distortion of a smectic-A liquid crystal mesophase in condensate droplets. These filamentous condensate structures share many similarities with myelin figures, which can instead form in surfactant mesophases [37-39]. From POM we validate that the filamentous condensate forms a liquid crystal phase with orientational ordering. Analysis of the birefringence pattern (Supplementary material section 1) indicates that within filaments the mesogens are aligned with homeotropic anchoring at the interface and distorted radially within the cylindrical interior with a defect line at their core (Figure 1C).

Spherical nuclei have a higher free energy density associated with mesophase splay distortion and are expected to spontaneously elongate into filaments with no energy barrier (Supplemental Materials section 2). Figure 2A shows the growth dynamics of a filament within a pore of the network at late times (Figure 2A; zoom from red dashed region in Figure 1B panel 3). Consistent with previous experimental observations [32, 34, 35] and theoretical treatments [30, 31, 33, 40], we observe filaments grow along their entire length $dl/dt \sim l$, resulting in exponential growth $l = l_0 \exp(kt)$ with a rate $k$ (Figure 2B). As suggested by previous theory [30], radial condensation of the mesogen into the surface of the filament results in a $k \sim R_F^{-2}$ scaling with filament radius, with $kR_F^2 \approx 0.27 \pm 0.01$ µm²/s at early times. At later times this growth is attenuated, likely due to local depletion of the condensing mesogen.

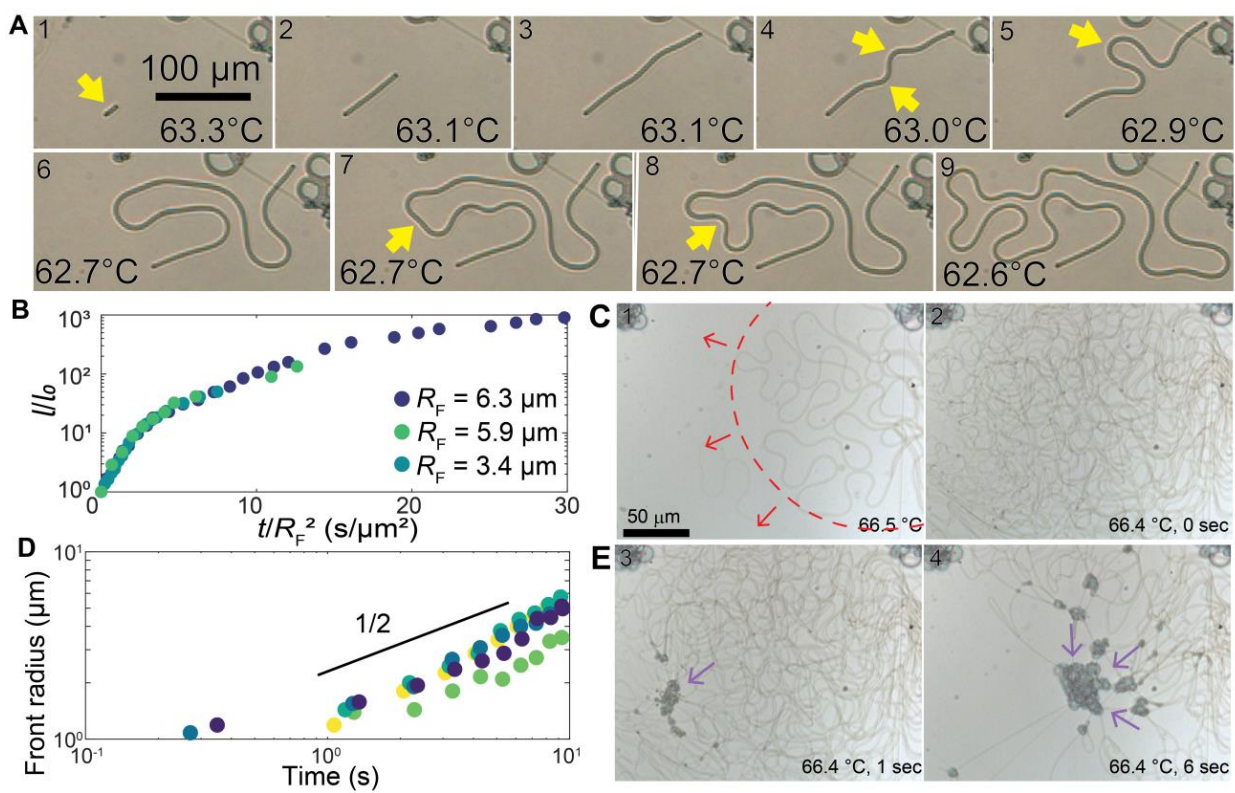

**Figure 2: Filament growth drives hydrodynamic buckling and dispersive densification. A** constant condensation along the entire filament length leads to **B** exponential increase in length. Hydrodynamics of growth lead to generational buckling. **C** Generational buckling leads to the formation of a dispersive front, **D** with a dispersion coefficient $D_* \approx 2.3 \pm 0.2$ µm²/s. **E** Above a critical density, filaments collapse into bulged discs (purple arrows), which aggregate in a rapid cascade (Movie S2).

Continued filament growth produces an extensile viscous flow, which is largest at the filament tips $U_{\text{tip}} \approx kl/2$. As a result, the viscous drag force experienced by the filament also grows $f_d = \int_0^l 8\pi c\mu U \, dx$, where $c$ is a shape factor arising from the slender body approximation and $\mu \approx$ 7 mPa·s is the viscosity of the squalane-rich phase [31]. We see this drag force lead to the eventual buckling of the filament (panels 3-4 of Figure 2A) consistent with a critical compressive force [33, 35]. For unbounded filaments, this generational buckling develops into a radially-dispersing filamentous front (dashed red line Figure 1B and Figure 2C). The radius of each filamentous front expands as $R_F \sim \sqrt{D_* t}$ (Figure 2C), from which we measure a dispersion coefficient $D_* \approx 2.3 \pm 0.2 \, \mu m^2/s$ (Figure 2D).

**Densified filaments collapse into bulged discs to relieve mesophase distortion**

The competition between exponential growth but quadratic dispersion produces a net densification over time $\rho \equiv l/R_F = \exp(kt)/\sqrt{D_* t}$. Above a critical filament density $\rho_c \approx 50$, we observe a rapid collapse of filaments into a dense aggregate (Movie S2; Figure 2E). Such aggregate structures have been observed previously, though their topological structure remains unclear [27-29, 32, 34]. POM imaging suggests these condensates form bulged discs, comprised of a wide cylindrical interior with no mesophase distortion surrounded by a toroidal periphery with radial splay mesophase distortion (Figure 1C). Comparison of the bulk distortion energy within the smectic mesophase further suggests that this bulged disc configuration has a substantially smaller volume-averaged free energy compared to the filamentous structure for large disc size (Supplemental Materials section 2). However, there is no barrier-free pathway for filaments to collapse into discs; instead, filaments seem to only pass the energy barrier to rearrange into bulged discs when they are sufficiently dense.

Once formed, discs dynamically retract connected filaments to relieve filament elastic distortion at velocities of $U_0 \approx 3$ to $30 \, \mu m/s$ (Movie S3; Figure 3A). Tangling of filaments due to both mechanical tension and transient viscous flows leads to a cascade of densification and collapse into smaller discs nearby, and mutual filament retraction rapidly brings these discs together to form a ramified aggregate. Indeed, we rarely observe discs in isolation. Discs within aggregates do not rapidly coalesce or coarsen, suggesting that another energy barrier to rearranging two adjacent toroidal boundaries to recombine into a larger disc structure.

**Viscous drag mediates retraction of filaments into discs**

The reduction in distortion energy between the filament and the undistorted disc interior suggests filaments are reeled in with a force of $f_r \approx \frac{\pi}{2} K \ln(R_F/R_C)$ (Supplemental Materials section 2), where $K$ is the bend modulus, and $R_C \approx 0.1$ nm is the defect core size, estimated to be on the order of the mesogenic molecule's size. The retraction of filaments at velocities in the range of $U_0 \approx 3$ to $30 \, \mu m/s$ suggest that this retraction force is overdamped by viscosity (low Reynolds numbers at a maximum $Re \approx \rho_F U_0 R_F/\mu \approx 10^{-5} \ll 1$, where $\rho_F$ is the mass density of the surrounding fluid). We can then predict the evolution of the length of a retracting filament by balancing the elastic retraction force with the viscous drag force (Supplementary Materials section 3), yielding the prediction:

$$(l/l_0)^2 = (l_c/l_0)^2 + (1 - (l_c/l_0)^2)\exp(kt),$$

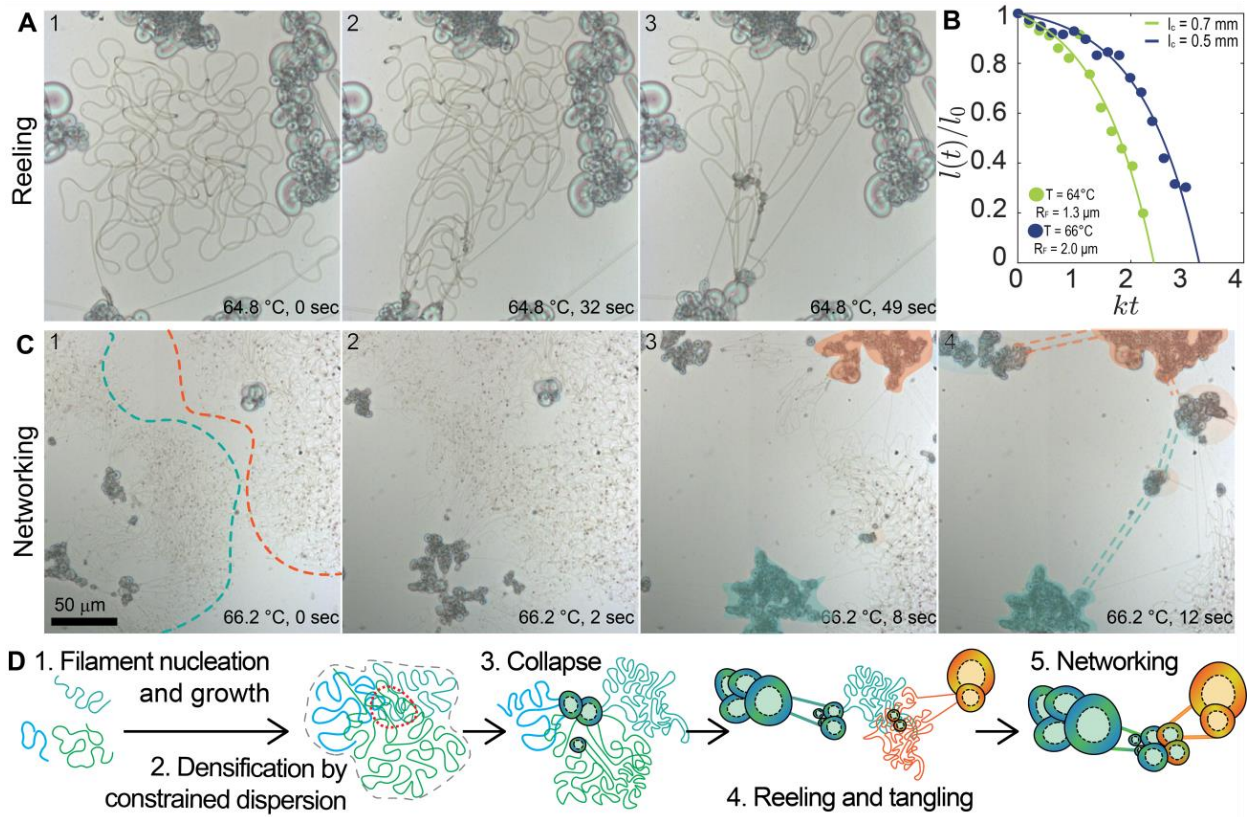

**Figure 3: Collapse of densified filaments into clusters of discs drives networking into porous scaffold.** **A** Discs reel in filaments to relieve splay deformation. **B** Short filaments experience an acceleration in shortening (Movie S3), consistent with theoretical predictions for $l_0 < l_c$ (lines show best fit for $l_c$). **C** In contrast, a filament above the critical length experiences net growth, leading to a dispersive corona which densifies and collapses into a new disc aggregate. Tangling of multiple filament coronas before collapse results in higher connectivity condensate network (Movie S4). **D** Schematic for hierarchical network formation driven by filament growth, densification, collapse, reeling, tangling, and networking.

which defines a critical filament length $l_c \equiv (K \ln(R_F/R_C)/(4\mu ck))^{1/2}$ below which the filament exhibits net shortening and above which the filament lengthens (Fig. S4). When $l_0 < l_c$, the filament shortening accelerates until termination, in good agreement with experimental tracking of individual retraction events (Figure 3B), giving estimates for $l_c \approx 0.5$ to $0.7$ mm.

Conversely, for a sufficiently long initial filament $l_0 > l_c$ the domination of growth leads to the formation of a filamentous corona around the disc aggregate, which continues to disperse. In this regime, unbounded growth once again competes with the dispersion of the filamentous corona, leading to net densification and collapse into additional aggregates of bulged discs. Filaments between aggregates are straightened by mutual retraction and remain under tension between aggregates; the aggregates remain separated. Successive iterations of this process lead to the formation of a sparse network of filaments under tension networked by disc aggregates. Tangling of multiple filamentous coronas prior to a collapse event (Figure 3C, Movie S4) provides higher connectivity networking that forms the eventual 2D porous scaffold (Figure 3D).

Thus, the formation of condensate networks follows a hierarchical assembly process. The initial condensation of this liquid crystal proceeds by nucleation and rapid growth of filaments, which are preferrable to spherical condensates because of reduced splay distortion energy of the smectic-A mesophase. Viscous drag leads to buckling and a dispersive confinement of the growing filamentous mass. The densifying filaments pass over an energy barrier to rearrange into bulged discs, with a lower free energy density. A rapid cascade of filamentous collapse leads to the clustering of dense disc aggregates, which retract filaments mediated by viscous drag and continued growth. Further expansion driven by the growth of filamentous coronas competes with contraction driven by elastic filament retraction, forming the macroscale sparse porous scaffold of filaments networked by disc aggregates. This image suggests that the architecture of the sparse porous scaffold can be tuned using the thermal ramp rate. Indeed, at successively higher thermal ramp rates up to 0.3 to 10°C/min the increased rate of collapse into disc clusters leads to a lower porosity network (Fig. S5), demonstrating the tunability across structural scales.

**Discussion**

Our work demonstrates that the phase separation of liquid crystals can produce rich nonequilibrium dynamics. The anisotropic growth of filaments converts the chemical potential of phase separation into microscale work that drives larger-scale hydrodynamic flows and macroscale pattern formation. This paradigm of flow generation is observed in a range of biological systems and leveraged in many biologically-inspired active fluids to drive spontaneous motion, in some cases producing large scale coherent flows or hierarchical material patterning [41-43]. This work is the first report, to our knowledge, that such nonequilibrium dynamics can be similarly harnessed in liquid crystal phase separation. Since liquid crystals arise broadly in nature and in biomacromolecules [44, 45], we anticipate these principles may arise in biological processes and can be leveraged for the biomimetic design of life-like materials.

Our observations further suggest a new route to assemble complex multi-level architectures by harnessing these nonequilibrium dynamics of liquid crystal phase separation. While dynamic external cues have been used to sequentially assemble multi-level nanomaterials [46], this work demonstrates how a single cue can pattern complex architectures. Since liquid crystals are widely used in optics and electronics [47, 48], filtration membranes [49-51], biological interfaces [52], microparticles [53], and metamaterials [26, 54], our approach may aid the self-assembly of industrially relevant materials and devices. Incorporation of promesogenic nanoparticles [55] or chromaphore dopants [56] into these filamentous and disc architectures may add further functionality for responsive materials and photon upconversion in optics and photovoltaics.

**Acknowledgments**. It is our pleasure to acknowledge Randall Kamien for insightful discussion. **Funding:** The authors gratefully acknowledge ExxonMobil for partial financial support of this work. This material is also based in part on work supported by the NSF Graduate Research Fellowship Program (to P.G.S.) under grant no. DGE-1845298. **Author contributions:** Conceptualization: M.G., E.S., Y.M., and C.O. Data curation: C.O. Formal analysis: Y.M., C.A.B., P.G.S., and C.O. Funding acquisition: M.G., E.S., C.O. Investigation: Y.M. and C.A.B. Methodology: Y.M., C.A.B., P.G.S., and C.O. Project administration: C.O. Resources: C.O. Software: Y.M. and Z.L. Supervision: C.O. Validation: Y.M., C.A.B., and C.O. Visualization: Y.M. and C.A.B. Writing-original draft: Y.M., C.A.B., and C.O. Writing-review & editing: C.A.B. M.G., E.S., O.A., K.E. and C.O. **Competing interests:** The authors have no competing interests to declare. **Data and materials availability:** All data needed to evaluate the conclusions in the paper are present in the paper and/or the Supplementary Materials.

**Author ORCIDs.**
Yuma Morimitsu 0000-0001-7303-0006;
Christopher A. Browne 0000-0002-3945-9906;
Zhe Liu 0000-0001-5017-7881;
Paul G. Severino 0000-0002-1648-4148;
Manesh Gopinadhan 0000-0001-8452-6613;
Eric B. Sirota 0009-0007-8328-7746;
Ozcan Altintas 0000-0003-3725-4613;
Kazem V. Edmond 0000-0001-6758-9110;
Chinedum Osuji 0000-0003-0261-3065;

## List of Supplementary Materials

Section 1: Materials and Methods
Section 2: Theoretical treatment of smectic distortion energy of filaments and discs
Section 3: Theoretical treatment of filament retraction into bulged discs
Supplementary Figures S1 to S5
Movie Captions S1 to S4

# Supplementary Materials for "Spontaneous assembly of condensate networks during the demixing of structured fluids"

**Section 1: Materials and Methods**

Our mesogenic liquid crystal is 4'-cyano 4-dodecyloxybiphenyl (*TCI Chemicals*), commonly referred to as 12OCB, which is known to form a smectic A liquid crystal mesophase in isolation. Our solvent is squalane (*TCI Chemicals*), which is a branched nonpolar alkane. All mixtures are prepared at high temperature >100°C, well above the transition temperature to the homogeneous isotropic phase.

We fabricate Hele-Shaw imaging cells using a glass slide and a coverslip, separated by dabs of a UV glue paste containing 60 μm silica beads, which sets the minimum spacing between slides. The slides are compressed and cured using UV illumination. The liquid crystal mixture is loaded into the Hele-Shaw cell at >100°C using capillary wicking, before the cell is fully sealed using UV glue. The final sample area is roughly 60 μm tall × 22 mm long × ~10 mm wide.

All imaging is conducted by placing the Hele-Shaw cell on a thermally controlled stage (*Linkam Scientific* THMS600), which is heated to 150°C before each test to ensure it has fully mixed before quenching into the binodal region. Subsequent cooling experiments do not exhibit any hysteretic system memory after reheating to 150°C. The thermal control stage and Hele-Shaw cell are inverted and placed on an inverted microscope (*Zeiss*) for imaging using brightfield (BF) and polarized optical microscopy (POM) at 5× or 20× magnification on 1000 × 1000 pixel square CCD (*Allied Vision Technologies* Pike).

Polarized optical microscopy (POM) of the condensing phase indicates the smectic A orientational structure, as schematized in figure 1C. Filaments exhibit a radial splay, indicated by the dark centerline and extinction when the filament is aligned with one of the polarizers. Aggregates are comprised of bulged discs, which have an undistorted interior that appears dark in POM since the director is aligned with optical axis. In the bulged periphery of the discs, the mesophase is radially splayed as indicated by extinction cross aligned with polarizers and decreasing fringe spacing as the bulged periphery gets thicker near the center disc.

**Section 2: Theoretical treatment of smectic distortion energy of filaments and discs**

Our toy model for the dynamics of the condensate network uses the relief of filaments' mesophase distortion as a driving force for filament retraction onto disks. Here we estimate the free energy associated with bulk distortion within the smectic-A filaments and bulged disks as seen in experiment. Typical bulk free energy functionals for smectic liquid crystals punish both compression and curvature of the smectic layers. Since both the filaments and bulged disks maintain perfect layer spacing, neglecting surface tension, we need only compare:

$$F_{\text{curv}} = \frac{K}{2} \int H^2 \mathrm{d}V,$$

where $H$ is the mean curvature of the smectic layers and $K$ is the bend modulus ($K$ can be related to the nematic splay coefficient $K_1 = K/4$ by noting that $H = \frac{1}{2}(\nabla \cdot \hat{\mathbf{n}})$). We first calculate the bulk free energy contribution of a filament as concentric cylinders of smectic layers, neglecting the ends, with maximum radius $R_F$ and length $l$. Each smectic layer has mean curvature $H = \frac{1}{2r}$, leading to $F_{\text{curv}} = \frac{\pi}{2} K l \ln\left(\frac{R_F}{R_C}\right)$, where $R_C$ is the defect core size, estimated to be on the order of the mesogenic molecule's size, $R_C \approx 0.1$ nm. While in general the secondary curvature induced from buckling must be considered for the behavior of the filaments themselves, we neglect this contribution when comparing between the filaments and disk configurations (note also that $R_F \ll R_{\text{buckle}}$ in our experiments). The volume-averaged curvature energy for the filaments is then:

$$\bar{F}_F = \frac{K}{2R_F^2} \ln\left(\frac{R_F}{R_C}\right),$$

We model the bulged disk configuration as flat, evenly-spaced smectic layers surrounded by toroidal periphery. Explicitly, the smectic layers of the periphery are concentric half tori, parameterized by

$$x = (R_b + R_s \cos(v)) \cos(u),$$
$$y = (R_b + R_s \cos(v)) \sin(u),$$
$$z = R_s \sin(v),$$

where $R_b$ and $R_s$ are the major and minor radii, $u \in [0, 2\pi)$ runs over the major axis, and $v \in [-\pi/2, \pi/2)$ runs over the minor axis. The curvature of each toroidal layer is then:

$$H = \frac{R_b + 2R_s \cos(v)}{2R_s(R_b + R_s \cos(v))}.$$

Since the smectic is in its ground state within the center of the bulged disk, we need only calculate the curvature energy from the periphery. Consider first the curvature energy of concentric tori (allowing $v$ to run from 0 to $2\pi$), rather than half tori. Integrating the mean curvature over the tori yields

$$\int_{V \in T^2} H^2 dV = \pi^2 R_b \left(-\tanh^{-1}\left(\sqrt{1 - \left(\frac{R_s}{R_b}\right)^2}\right) + \ln\left(\frac{R_b}{R_C}\right) + \ln(2)\right).$$

Although the "inside" and "outside" half of the torus have different mean curvatures, in the limit $R_b \gg R_s$ the free energy of the half torus periphery is half of that of the full concentric tori. Note also that the outside half of the torus necessarily has larger mean curvature than the inside; even in the case that $R_b$ is only somewhat larger than $R_s$, this expression overestimates the curvature energy of the disk periphery. This leads us to a bulk free energy per unit volume of

$$\bar{F}_D = \frac{K\pi}{2}\left(\frac{R_b}{2\pi R_s(R_b - R_s)^2 + \pi^2 R_s^2 R_b}\right)\left(-\tanh^{-1}\left(\sqrt{1 - \left(\frac{R_s}{R_b}\right)^2}\right) + \ln\left(\frac{R_b}{R_C}\right) + \ln(2)\right)$$

for the entire disk configuration. Experimentally we typically observe $R_b > 10 R_s$, and for analytic simplicity we analyze the expression for $\bar{F}_D$ in the limit $R_b \gg R_s$:

$$\bar{F}_D \approx \frac{\pi K}{4 R_s R_b} \ln\left(\frac{R_s}{R_C}\right),$$

which demonstrates why the bulk free energy of the disk state is preferable to filaments. The disk configuration has the same logarithmic contribution to the free energy as the filaments, corresponding to the `half filament' toroidal periphery required for homeotropic anchoring. In contrast to the filaments in which all smectic layers have curvature distortion, the bulged disk configuration allows smectic layers in the disc interior to remain in their ground state. Taking the ratio between the free energy densities of a filament (of any length) and a bulged disc with $R_s = R_F$, we expect the bulged disc configuration to be favorable when:

$$R_b > \frac{\pi R_F}{2},$$

suggesting a threshold size above which bulged discs will grow and retract any connected filaments. This free energy further suggests that bulged disks grow primarily by expanding their major radius, consistent with experimental observations.

**Section 3: Theoretical treatment of filament retraction into bulged discs**

We consider a filament that retracts from one end at velocity $U_0(t) \equiv U(t, x = 0)$ as it continues to grow along the remainder of its length $x = [0, \ l]$. This is identical to a filament of length $2l$ retracting at both ends with the same reeling velocity at each end. A mass balance along the growing filament yields:

$$\frac{dU}{dx} = -k,$$

with the boundary condition $U(t, x = 0) \equiv U_0(t)$ giving the velocity profile:

$$U(x) = U_0 - kx.$$

Approximating the viscous drag using the slender body approximation gives the force along the filament $f_d$:

$$f_d = \int_0^l 8\pi c \mu U(x) dx.$$

The slender body approximation introduces a shape factor $c$ accounting for the aspect ratio (shelly). In principle, $c \sim 1/\ln(l/R_F)$, though we consider it to be a constant to allow for analytic integration.

The movement of material from the retracting filament into the growing bulged disc transfers a differential volume $\Delta V = \pi R_F^2 \Delta x$, where $\pi R_F^2$ is the cross sectional area and $\Delta E$ is a differential translation of the filament in the vicinity of the bulged disc, such that $\lim_{\Delta t \to 0} \Delta x/\Delta t = U_0$. The change in free energy density associated with this of the translating filament and reduces the internal energy of the system by a differential amount:

$$\Delta E = \Delta V (\bar{F}_F - \bar{F}_D),$$

which can be reinterpreted as a force to retract the filament $f_r = \Delta E/\Delta x$:

$$f_r = \pi R_F^2 (\bar{F}_F - \bar{F}_D).$$

Inserting our expression for $\bar{F}_D$ in the limit $R_b \gg R_s$:

$$f_r \approx \frac{\pi K}{2} \ln\left(\frac{R_F}{R_C}\right)\left(1 - \frac{\pi R_F^2 \ln\left(\frac{R_s}{R_C}\right)}{2 R_s R_b \ln\left(\frac{R_F}{R_C}\right)}\right),$$

which in the limit $R_s > R_F$ and $R_b \gg R_F$ further simplifies to:

$$f_r \approx \frac{\pi K}{2} \ln\left(\frac{R_F}{R_C}\right),$$

indicating that the free energy density of the bulged disc is sufficiently smaller than the filament, such that it does not strongly affect the retraction force. Since the system is overdamped by viscosity ($Re < 10^{-5}$), we expect the retraction force to balance the drag force, with minimal inertial effects associated with acceleration:

$$f_r \approx f_d,$$

$$\frac{\pi K}{2} \ln\left(\frac{R_F}{R_C}\right) \approx 8\pi c \mu \int_0^l U(x) dx.$$

We then insert the velocity profile $U(x) = U_0 - kx$ to integrate:

$$\frac{K \ln\frac{R_F}{R_C}}{8\mu c} = U_0(t) l(t) - \frac{k}{2} l^2(t).$$

Finally, we insert the macro-scale mass balance:

$$\frac{dl}{dt} = kl - U_0.$$

Rearranging for $U_0$ and substituting into the force balance gives the differential equation:

$$\frac{d(l^2(t))}{d(kt)} = l^2(t) - \frac{K \ln\frac{R_F}{R_C}}{4\mu c k}.$$

Stability of this differential equation suggests the filament should experience net retraction when:

$$l^2(t) < \frac{K \ln\frac{R_F}{R_C}}{4\mu c k} \equiv l_c^2,$$

which can be interpreted as a critical filament length. Integrating from some initial length $l(t=0) \equiv l_0$ gives:

$$\left(\frac{l(t)}{l_0}\right)^2 = \left(\frac{l_c}{l_0}\right)^2 + \left(1 - \left(\frac{l_c}{l_0}\right)^2\right) e^{kt},$$

where $l_0/l_c$ could also be interpreted as a nondimensional quantity comparing the length to a critical length or equivalently as the relative rates of growth and retraction. Lines in figure 3C show best fit $l_c$ for each filament, which are observed at different temperatures $T$ and filament radii $R_F$.

# Supplementary Figures

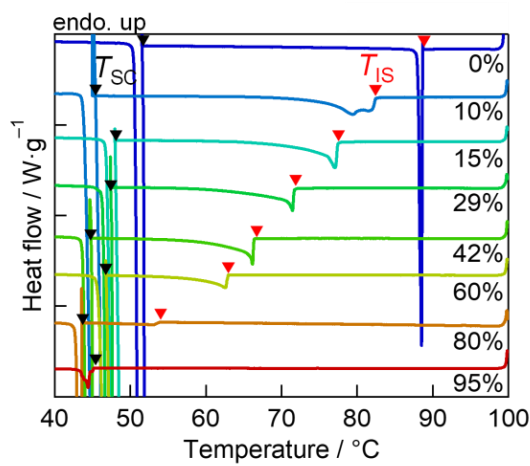

**Fig. S1. Differential Scanning Calorimetry (DSC).** DSC is conducted using a *TA DSC2500* for various compositions of mesogen (4'-cyano 4-dodecyloxybiphenyl, *TCI Chemicals*) and solvent (squalane, *TCI Chemicals*) cooled from 100 ˚C (where all solutions are isotropic liquids) at ramp rate of 1˚C/min. Percentages displayed in figure indicate the weight percent of squalane. Red and black triangles indicate the transition temperatures from isotropic to smectic A (labeled $T_{IS}$) and from smectic A to crystal (labeled $T_{SC}$), respectively.

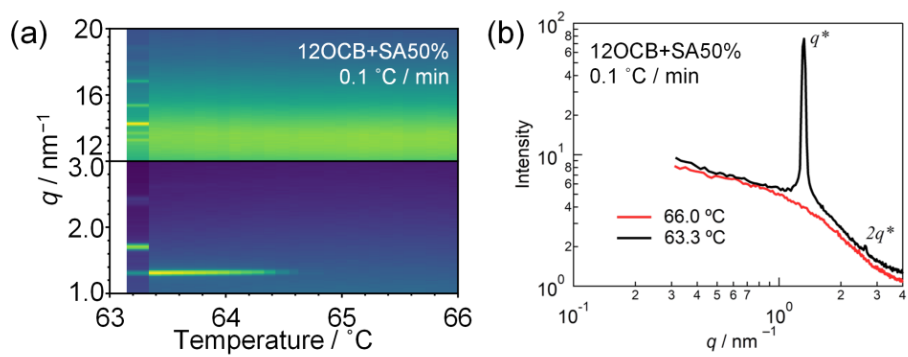

**Fig. S2. Small Angle X-Ray Scattering (SAXS) and Wide Angle X-Ray Scattering (WAXS) profiles.** (a) 50 wt.% mesogen in squalane solvent is raised to 150˚C to homogenize well above the isotropic transition and then rapidly brought to 70˚C. SAXS and WAXS profiles are measured during continuous cooling at a rate of 0.1˚C/min. Stability of this peak indicates the nucleation of smectic A mesophase is not a transient metastable state. (b) Selected SAXS profile at temperatures of 66.0 and 63.3 ˚C indicate that the fluid is isotropic at high temperatures and smectic A below the transition observed in DSC. The smectic layer spacing is $h \approx 0.8$ nm.

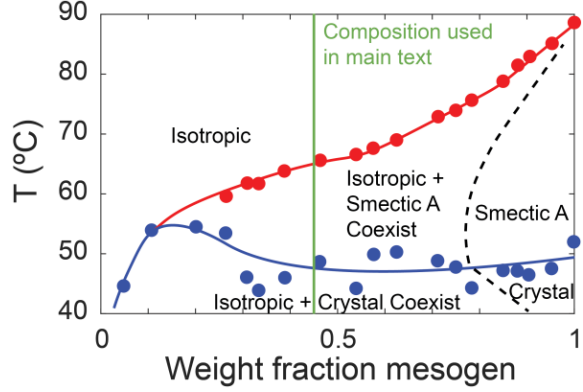

**Fig. S3. Phase diagram.** Red points indicate transition from isotropic to isotropic-smectic A coexistence, as determined by DSC and SAXS/WAXS. Blue points indicate transition from isotropic-smectic A coexistence to isotropic-crystal coexistence. Red and blue lines are to guide the eye. Black dashed lines are presumed boundaries, sketched using solubility measurements (not shown). All experiments in the main text are conducted at 0.45 weight fraction of mesogen, as indicated by green line.

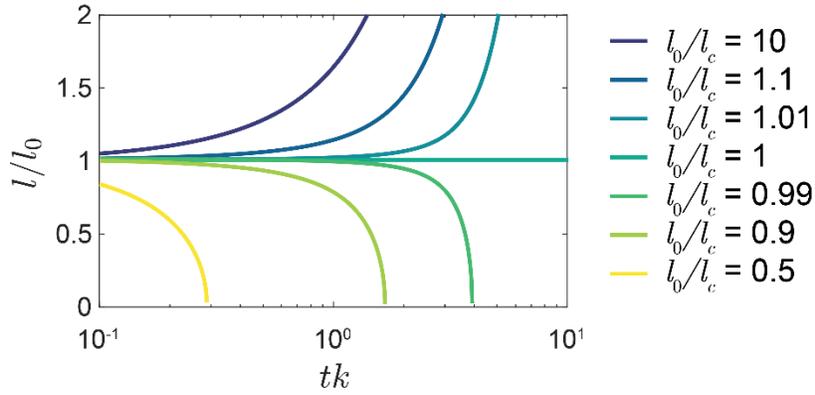

**Figure S4. Filament retraction theory.** Theory for simultaneous growth and retraction of a filament into bulged disc aggregate at a range of $l_0/l_c$. For $l_0/l_c > 1$ the filament exhibits net growth, while at $l_0/l_c < 1$ the filament experiences net retraction.

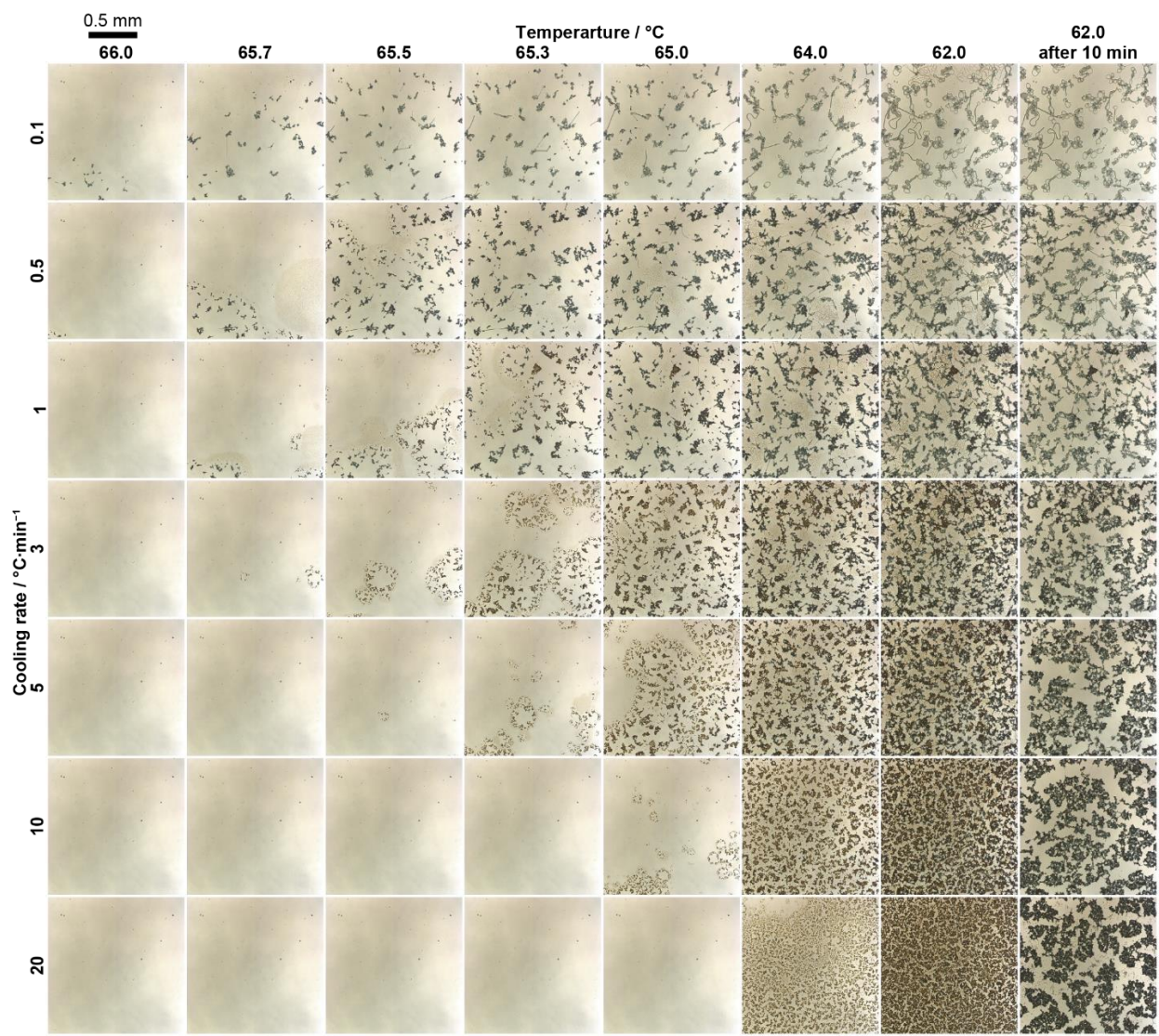

**Fig. S5. Phase separation at various cooling rates.** Sample is 45 wt.% mesogen and balance solvent. Images obtained using brightfield microscopy under 5× magnification and recorded on a 1000×1000 pixel CCD camera (*Allied Vision Technologies* Pike). Smectic-A condensate phase appear as dark spots due to increased optical scattering. Images in Figure 1 of main text are false colored using thresholding.

# Movie Captions

Access here: https://upenn.box.com/s/iepjtvwd6o5ur8wkuiphxjigyaljckiv

**Movie S1:** Brightfield (left) recording of phase separation from 70°C to 62°C cooling at a constant rate of –0.1°C/min. Video taken on 5× magnification. Field of view is 1.4 mm × 1.4 mm, shown at 60× real time. Right is post-processed overlay of filaments (false colored blue) and bulged disc aggregates (yellow), identified using image thresholding.

**Movie S2:** Brightfield recording of filament growth, dispersion, densification, and collapse into bulged disc aggregate. Video taken on 20× magnification. Field of view is 357 μm × 357 μm. Shown at 15× real time.

**Movie S3:** Brightfield recording of filament reeling into bulged disc aggregates. Video taken on 20× magnification. Field of view is 357 μm × 357 μm. Shown at 15× real time.

**Movie S4:** Brightfield recording of filament tangling and collapse into networked aggregates of bulged discs. Video taken on 20× magnification. Field of view is 357 μm × 357 μm. Shown at 15× real time.